\documentclass[useAMS,usenatbib,usegraphicx]{mn2e}
\usepackage{graphicx,amssymb,amsmath}
\usepackage{color}

\title[Blazhko light curves]{Fitting Blazhko light curves}
\author[Szeidl et al.]{B. Szeidl$^{1}$, J. Jurcsik$^{1}$, \'A. S\'odor$^{1}$, G. Hajdu$^{1}$, P. Smitola$^{2}$
\\
$^{1}$Konkoly Observatory of the Hungarian Academy of Sciences, H-1525 Budapest PO Box 67, Hungary\\
$^{2}$E\"otv\"os University, Dept. of Astronomy, H-1518 Budapest PO Box 49, Hungary}
\begin{document}

\date{Accepted 2012 June 7,  Received 2012 June 1; in original form 2012 May 2}

\pagerange{\pageref{firstpage}--\pageref{lastpage}} \pubyear{2012}

\maketitle

\label{firstpage}
\begin{abstract}
The correct amplitude and phase modulation formalism of the Blazhko modulation is given. The harmonic order dependent amplitude and phase modulation form is equivalent with the Fourier decomposition of multiplets.
The amplitude and phase modulation formalism used in electronic transmission technique as introduced by Benk\H o, Szab\'o and Papar\'o (2011, MNRAS 417, 974) for Blazhko stars oversimplifies the amplitude and 
phase modulation functions thus it does not describe the light variation in full detail.

The results of the different formalisms are compared and documented by fitting the light curve of a real Blazhko star, CM UMa.

\end{abstract}

\section{Introduction}
The periodic modulation of the light curve of a large percentage of RR Lyrae stars, the so-called Blazhko effect, is a hundred-year old enigma. The height and time of maximum light of these stars oscillate with the same period of several days, weeks, months or even years. \citet{bl} was the first who noticed that no constant period could satisfy the observed times of maximum light of RW Dra, an RR Lyrae type star\footnote{At the beginning of the $20^{\mathrm {th}}$ century the origin of the light variation of RR Lyrae (cluster-type) variables was unknown and they were classified according to the shape of their light curves as `antalgol' type.}, and an oscillation in the fundamental period with a cycle length of 41.6 days had to be postulated. The striking changes in the height of the light maximum and in the shape of the light curve of RR Lyr, which turned out to be periodic with 40 days, was discovered by \cite{sh}. Following these discoveries, the Blazhko effect has been detected and studied in a number of RR Lyrae stars.

The large-scale surveys as MACHO and OGLE \citep{alcock,scz03,scz11} detected 10\,--\,30 per cent incidence rate of the modulation among RRab stars in the Magellanic Clouds and the Galactic bulge, while  
a recent ground-based multicolour photometric survey \citep{kbs} and
the highly accurate observations of the CoRoT and {\it Kepler} space missions \citep{ch09,k10} revealed that a significantly larger fraction of the fundamental-mode RR Lyrae stars (about 50 per cent) shows the effect.

The high occurrence rate of modulated RR Lyrae stars makes the Blazhko effect an even more intriguing problem of the pulsation theory. No wonder that it has captivated the researchers' interest again in recent years. In spite of the fact that the highly accurate space data and the ground-based multicolour photometries have given new insights into the Blazhko phenomenon \citep[for a recent review see][]{k11}, no generally accepted theory exists that is able to explain the observed features of the effect \citep{ko09}.

The connection between the shock-waves in the atmosphere and light-curve modulation is particularly interesting. The shock-waves play a significant role in the appearance of the bump/hump features on the light curve of RR Lyrae stars. \cite{psp}, investigating the spectral-feature variations during the 41-day modulation cycle of RR Lyrae, suggested a model in which `a critical level of shock-wave formation moves up and down' in the star's atmosphere. Later,
\cite{cg97} have found that the amplitude of shock-waves is greatly correlated with the Blazhko modulation.  

To describe the light variations of Blazhko stars, mathematical models have been propounded. A model that aimed to describe the full Fourier spectrum of the modulation was suggested by \cite{bk}. The triplet was interpreted as the non-linear coupling of the main pulsation mode and a non-radial mode close to it, and many of their combination terms. However, no detailed confrontation of the possible predictions of this model with the observations (e.g. on the amplitudes of the components of the multiplets) has been performed. In other approaches \citep[][hereafter BSP11]{szj,b11} the multiplet structure is the simple result of the modulation of purely radial pulsation. 

In their comprehensive study, BSP11 presented an analytical formalism (applied in electronic signal transmission) for the description of the light curves of Blazhko RR Lyrae stars. Their
model shows several light-curve characteristics similar to 
those observed in real Blazhko stars. It was also claimed that the new light-curve solution drastically reduced the number of necessary parameters compared to the traditional methods. However, in BSP11, the suggested method was not tested on real observational data.

Recently, \cite{gug} made an attempt at applying the proposed new analytic modulation formalism to the Kepler data of the complex Blazhko star KIC 6186029 = V445 Lyr.
Although the model light curve showed the global properties of the observed one,
the fitted light curve deviated from the observed data significantly in certain phases of the
pulsation and the modulation \citep[see figure 16 in][]{gug}.
The surprisingly high variance of the residuals was explained by method-specific and
object-specific reasons. The method did not describe the migration of the bumps and humps,
and as any `stationary' model, it was unable to follow an irregular, time-dependent phenomenon \citep{gug}. 

In this paper we look into the possible short-comings of the formalism introduced in BSP11. The
mathematical formalisms are given in Sect. 2, and the results are documented by the different
fits of the light curve of a Blazhko star, CM UMa in Sect. 3. As the modulation of CM UMa is
quite simple and regular; no secondary modulation is detected, and no higher order than quintuplet
modulation components are present in the Fourier spectrum, therefore no bias of the results arises from any time-dependent irregularity of the modulation, what was the case in V445 Lyr.

\section{Mathematical considerations}

The Fourier decomposition of the light curve of Blazhko RR~Lyrae stars is a common technique to analyse their modulation properties. If $f_0$ and $f_\mathrm{m}$ indicate the fundamental and modulation frequencies, in the typical Fourier representation of the light curve, the harmonics ($if_0$) and the multiplets (triplets, quintuplets, septuplets, etc.: $if_0\pm jf_\mathrm{m}$), as well as the harmonics of the Blazhko frequency ($kf_\mathrm{m}$) appear in the spectrum.

Let  $\omega=2\pi f_0$ and $\Omega = 2\pi f_\mathrm{m}$, then the Fourier representation of the modulated light curve is:

\begin{multline}\label{eq:four} 
m(t) = m_0 + \sum_{k=1}^l b_{k} \sin \left( k \Omega t + {{\varphi}_{\mathrm{b}k}}\right) +\\ \sum_{i=1}^n \bigg[ a_{i} \sin \left( i\omega t + \varphi_{i} \right) + \sum_{j=1}^{l_i^+} a_{ij}^+ \sin \left( i\omega t + j\Omega t +\varphi_{ij}^+\right)+\\
\sum_{j'=1}^{l_i^-} a_{ij'}^- \sin \left( i\omega t - j'\Omega t + \varphi_{ij'}^-\right)\bigg]
\end{multline}

\noindent where the amplitudes ($b, a$) and angles (${{\varphi}_\mathrm{b}}, \varphi$) are constants (latter ones are epoch dependent).
% and the $k$,$i$ and $j$,$j'$ indices denote the harmonic orders of the Fourier series of the modulation and the pulsation, and the multiplet orders of the `$+$' and `$-$' side modulation components, respectively. 
In the expression of $m(t)$, the first sum $\sum_{k=1}^l b_k \sin \left( k \Omega t + \varphi_{\mathrm{b}k}\right)$ corresponds to the mean light-curve variation during the Blazhko cycle.

The application of  formula \ref{eq:soksin} (see Appendix A) to Eq.~\ref{eq:four}  leads to the following form:

\begin{multline}
\label{eq:summod}
m(t) =  m_0 + \sum_{k=1}^l b_{k} \sin \left( k\Omega t + {\varphi}_{\mathrm{b}k} \right)+\\
\sum_{i=1}^{n}[a_{i}+ f_{\mathrm{A}i}(t)] \sin \left[ i\omega t +{\varphi_{i}} +f_{\mathrm{F} i}(t) \right] 
\end{multline}

\noindent where $a$, $b$, $\varphi$ and ${\varphi}_\mathrm{b}$ are amplitude and phase constants, while $f_{\mathrm{A}i}(t)$ and $f_{\mathrm{F}i}(t)$ are the amplitude and angle (phase or frequency, see Appendix B) modulation functions of the $i$th harmonics of the pulsation.
They depend only on $\Omega(t)=2\pi f_\mathrm{m} t$ and the constant parameters of the Fourier series Eq.~\ref{eq:four}. Therefore, they are periodic functions according to $f_{\mathrm{m}}^{-1}$ ($=2\pi\Omega^{-1}$), consequently, they can be approximated by the Fourier series:

\begin{equation}\label{amod}
f_{\mathrm{A}i}(t) = \sum_{j=1}^{l_{i}^\mathrm{A}}  a^\mathrm{A}_{ij}  \sin \left( j\Omega t + \varphi_{ij}^\mathrm{A} \right)
\end{equation}

\noindent and

\begin{equation}\label{pmod}
f_{\mathrm{F}i}(t) = \sum_{j'=1}^{l_{i}^\mathrm{F}} a^\mathrm{F}_{ij'} \sin \left( j'\Omega t + \varphi_{ij'}^\mathrm{F} \right),
\end{equation}

\noindent where $a^\mathrm{A}$, $a^\mathrm{F}$, $\varphi^\mathrm{A}$ and $\varphi^\mathrm{F}$ are amplitude and phase constants.

%Here the $j$,$j'$ indices refer to the harmonic orders of the amplitude and phase modulation series.
In Eqs.~\ref{eq:four}, \ref{eq:summod}, \ref{amod} and \ref{pmod}, the harmonic orders of the pulsation and the different modulation series
$n,l,l_i^+,l_i^-, l_i^\mathrm{A}, l_i^\mathrm{F}$ are limited by the accuracy of the observations. 

The equality of Eqs.~\ref{eq:four} and \ref{eq:summod} demands that the time-history of a modulated star must be equally well described by either of them, with about the same variance of the residual, no matter which of the two formalisms is chosen. Then, the question is raised why does the formalism of BSP11 give an inferior fit to the Fourier decomposition? If Eq.~\ref{eq:summod} is compared to Eq.~49 of BSP11, the difference is at once conspicuous. In Eq.~\ref{eq:summod}, the amplitude and angle modulation functions (Eqs.~\ref{amod} and \ref{pmod}) depend on the harmonic order ($i$) of the Fourier sequence of the unmodulated light curve (carrier wave), whereas BSP11 assume that the modulation functions depend on the harmonic order in a very restricted way. To be more specific, in their approach, each harmonic term of the Fourier series of the unmodulated light curve is modulated in amplitude and angle by the   

\begin{equation}
\label{bamod}
g_{\mathrm{A}}(t) = a^{\mathrm A}_0 + \sum_{p'=1}^{q'} a^{\mathrm A}_{p'} \sin\left( 2\pi p' f_{\mathrm m} t + \varphi^{\mathrm A}_{p'} \right)
\end{equation}

\noindent and

\begin{equation}\label{bpmod}
g_{{\mathrm{F}}{i}}(t) =  i \sum_{p=1}^{q} a^{\mathrm F}_p \sin \left( 2\pi p f_{\mathrm m} t + \varphi^{\mathrm F}_p \right)
\end{equation}

\noindent formulae, respectively. 
We note here that in BSP11, the constant term of the amplitude modulation function $g_\mathrm{A}(t)$ is identified with the difference between the magnitude and intensity mean values,
and the constant term of the phase modulation function $g_\mathrm{F}(t)$ is contracted into the $\varphi_i$ phase constant of the pulsation.

In Eqs.~\ref{bamod} and \ref{bpmod}, the BSP11 notation holds, with the exception of denoting the harmonic order of the pulsation with $i$, instead of $j$. Here $g_{\mathrm{A}}$ does not depend on the harmonic order of the unmodulated light curve in the least, which is inconsistent with Eq.~\ref{amod}, while in $g_{\mathrm{F}_i}$, the harmonic order is not taken into account correctly with the sole multiplication of the modulation function by $i$ (cf. Eq.~\ref{pmod}).

\subsection{Consequences}

Due to the adoption of simplified formulae for the amplitude and angle modulations, BSP11 drastically reduce the number of parameters in their formalism. However, the outcome is an inferior fit compared to the Fourier fit (see details in Sect.~\ref{case}). The modulation assumed by BSP11 can be effectuated in electronic signal transmission, but, a priori, it is not known  how the star modulates its pulsation light curve (the `signal').

Another serious problem is the change of the mean brightness during the Blazhko cycle. Obviously, the angle modulation cannot contribute to it. BSP11 state that `the formalism naturally explains the mean brightness variations',
however, in their description it is directly proportional to the amplitude modulation. Like their amplitude modulation formalism describes only an averaged harmonic order dependence, it can happen that the mean varies in a more complex way than the BSP11 formalism demands. In real stars, it is actually observed: e.g., neither the magnitude nor the intensity averaged variation of the mean $V$ brightness of MW Lyr is linearly proportional to the amplitude variation of the star during the Blazhko cycle \citep[see figure 14 in][]{mw2}. Moreover, their difference is Blazhko phase dependent, thus it cannot be taken into account as a sole constant, as in BSP11.

In principle, the summation of $j$ and $j'$ in Eqs.~\ref{amod} and \ref{pmod}  can start from zero; the $j=j'=0$ terms modify the amplitude and phase of the unmodulated light curve in each harmonic order. These constant terms are included in the  $a_i$ and $\varphi_i$ parameters of Eq.~\ref{eq:summod}. Therefore, 
the choice of the starting index of the summation from 1 does not have any consequence on the results.
These extra terms express the differences between the mean light curve of a Blazhko star and the light curve of an unmodulated RRab star with the same physical properties. 

%The existence or non-existence of these extra terms might be an important contribution to the modulation properties of Blazhko stars, as they indicate  As this is an unsolved problem, 

The difference between the phase and amplitude modulation functions of BSP11 and the present paper (modulation functions deduced from the Fourier series) has another consequence, too. The observations prove that the amplitude versus phase diagrams for the different harmonic orders of the light curve during the Blazhko cycle show intrinsically different forms (see details in Sect.~\ref{a-p}). These diagrams determined according to the formalism of BSP11 are morphologically exactly the same for each harmonic order. This fact also refers to that the reduction of the degree of freedom (decreasing the number of free parameters) by BSP11 alters the real phase and amplitude relations of the observations.

A further important question, which is generally regarded as trivial, is the definition/determination of the phase difference between the angle and the amplitude modulations. In Eqs.~\ref{amod} and \ref{pmod}, the angles depend on the starting epoch, but their difference $\Phi_{ij}={\varphi_{ij}^A}-{\varphi_{ij}^F}$ is epoch independent. Since the $\Phi_{ij}$ values can be different for the possible combinations of $i$ and $j$, the definition of the phase difference between the angle and the amplitude modulations is not at all obvious. %This problem deserves further study (kell ez?)

\section{A real case study: CM~UMa}
\label{cmu}
\begin{figure}
\begin{centering}
\includegraphics[width=9.2 cm]{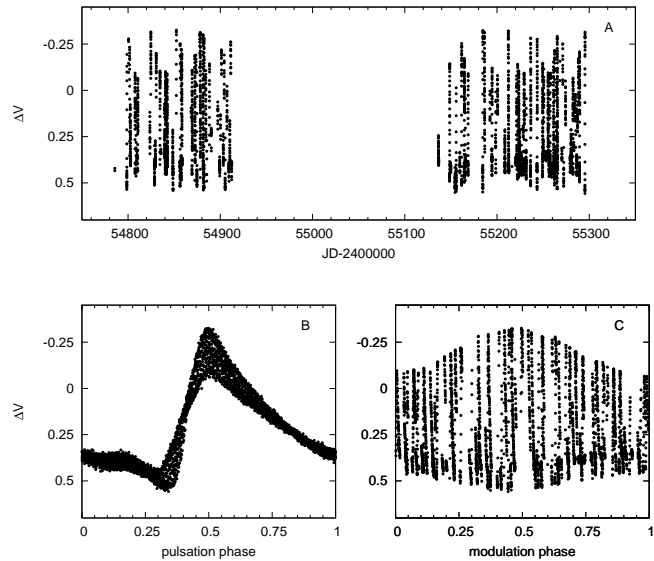}
\caption{The $V$ light curve of CM~UMa and the light curve phased with the pulsation (0.589124 d) and modulation (27.77 d) periods are shown in panels A, B and C, respectively.}
\label{lc}
\end{centering}
\end{figure}

In this section, we compare the results of fitting the light curve of a real Blazhko star with \\
-- the Fourier series of Blazhko-frequency-spaced multiplets around each pulsation frequency component (Case~A,~Eq.~\ref{eq:four}); \\
-- the amplitude and phase modulated sum of the pulsation light curve (Case B, Eq.~\ref{eq:summod});\\
-- the formalism introduced by BSP11 [Eq.~49 therein] (Case C, Eq.~\ref{eq:benko}).

In order to  index and parametrize the equations in a homogeneous manner,  Eq.~49 of BSP11 is rewritten in the following form:

\begin{multline}\label{eq:benko}
m(t)= {m_0}' + \left[ a^{\mathrm A}_0+ \sum_{j=1}^{l^\mathrm{A}} a^{\mathrm A}_{j} \sin \left( j \Omega t + \varphi^{\mathrm A}_{j} \right) \right] \cdot \\ 
\left\{ a_{0} +\sum_{i=1}^n a_i \sin \left[  i \omega t +  
i \sum_{j'=1}^{l^\mathrm{F}} a^{\mathrm F}_{j'} \sin \left(  j' \Omega t + \varphi^{\mathrm F}_{j'} \right) + \phi_i \right] \right\}.
\end{multline}
\noindent
The only difference between Eq.~49 of BSP11 and Eq.~\ref{eq:benko} is an additional term: ${m_0}'$. Without adding this extra constant to the equation, a real light curve cannot be fitted.  We also note that the magnitude zeropoint of Eq.~\ref{eq:benko} is not identical with the zero points of Eqs.~\ref{eq:four} and \ref{eq:summod}: ${m_0}'= m_0-a^{\mathrm A}_0 \times a_{0}$.

CM~UMa, one of the Blazhko stars,  which was extensively observed in the course of the Konkoly Blazhko Survey~II \citep{kbs2} has been chosen for the case study.  CM~UMa was observed with the 60-cm automatic telescope of the Konkoly Observatory (Budapest, Sv\'abhegy) in the 2009 and 2010 observing seasons.  The detailed analysis of the multicolour light curve of CM~UMa will be published elsewhere.

The light curve of CM~UMa can be described simply with the combination terms of one pulsation ($f_0$) and one modulation ($f_\mathrm{m}$) frequency components.  No secondary modulation or additional frequency has been detected in the Fourier spectrum, which makes CM~UMa an ideal target for such an investigation. Its light-curve solution consists of the Fourier sum of quintuplets around the pulsation frequency and its harmonics, and the $f_\mathrm{m}$ and $2f_\mathrm{m}$ modulation frequencies.

The $V$ light curve of CM~UMa (shown in Fig.~\ref{lc}) is fitted with different order series of the pulsation ($n$) and the modulation ($l,{l_i^+},{l_i^-}{l_i^\mathrm{A}},{l_i^\mathrm{F}},l^\mathrm{A},l^\mathrm{F}$) components according to Eqs.~\ref{eq:four}, \ref{eq:summod} and \ref{eq:benko}. In each solution, the harmonic orders of the `$+$' and `$-$'  side multiplets or the amplitude and phase modulations are taken to be identical and the same for each harmonic of the pulsation, i.e. $N={l_i^+}={l_i^-}={l_i^\mathrm{A}}={l_i^\mathrm{F}};\,\,\, i=1,..n$ (Eqs.~\ref{eq:four}, \ref{amod} and \ref{pmod}). When applying Eq.~\ref{eq:benko}, the orders of the amplitude- and phase-modulation series  are also taken to be identical: $N=l^\mathrm{A}=l^\mathrm{F}$. 
The zero point of the magnitudes ($m_0$, ${m_0}'$) are also fitted in all the solutions. A low-order ($l=1,2$) sum of the modulation frequency in Case A and B has been taken into account, too. The fits thus comprise
$p=n(4N+2)+2l+1$ parameters when using Eqs.~\ref{eq:four} and \ref{eq:summod}, while $p=2n+4N+3$  for Eq.~\ref{eq:benko}.

The pulsation and modulation frequencies are not fitted in any of the procedures;
$f_0=1.697434$ and $f_\mathrm{m}=0.036006$, determined from the Fourier analysis of the light curve are accepted and taken as constants. 

\subsection{Comparison of the accuracies of the fits}
\label{case}
\begin{table*}
\centering
\caption{Comparison of the rms of the residuals when fitting the light curve of CM~UMa according to Eqs.~\ref{eq:four}, \ref{eq:summod} and \ref{eq:benko}.\label{rms}
}
\label{residual}
\begin{tabular}{cccccc}
\hline
\multicolumn{2}{c}{Case A (Eq.~\ref{eq:four})}  &\multicolumn{2}{c}{Case B (Eq.~\ref{eq:summod}) }& \multicolumn{2}{c}{Case C (Eq.~\ref{eq:benko})}\\ 
$p(n,N,l)$&rms [mag]&$p(n,N,l)$&rms [mag]&$p(n,N)$&rms [mag]\\
\hline
185(18,2,2) & 0.00911& 185(18,2,2)&0.00928& &\\
135(13,2,2) & 0.00945& 135(13,2,2)&0.00976& &\\
\,83(13,1,2)& 0.00987&\,83(13,1,2)&0.01033&63(20,5)&0.01924\\
\,57(9,1,1) & 0.01056&\,57(9,1,1) &0.01100&55(18,4)&0.01930\\
\,39(6,1,1) & 0.01213&\,39(6,1,1) &0.01254&43(16,2)&0.01946\\
\hline
\end{tabular}
\end{table*}

\begin{figure}
%\begin{centering}
\includegraphics[width=8.8 cm]{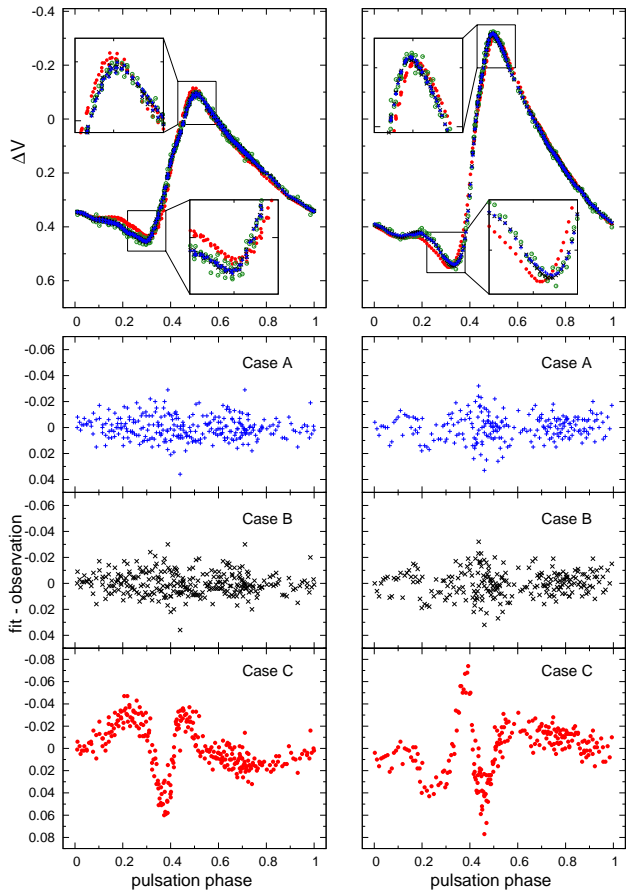}
\begin{centering}
\caption{The light curves of CM~UMa  in the  lowest and highest amplitude phases of the Blazhko modulation (empty circles) and the fits according to Eqs.~ \ref{eq:four} ($p$=185),  \ref{eq:summod} ($p$=185) , and \ref{eq:benko} ($p$=55) are denoted by $+$, x and $\bullet$ symbols, respectively (top panels). The inserts magnify the light curves and the fits at minimum and maximum light. The bottom panels show the residuals of the fits.}
\label{fit}
\end{centering}
\end{figure}

\begin{figure}
%\begin{centering}
\includegraphics[width=8.8 cm]{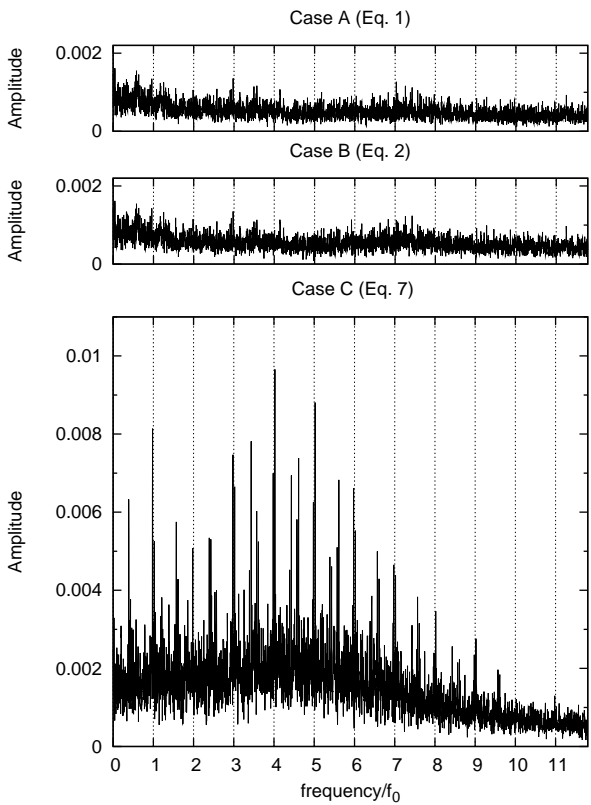}
\begin{centering}
\caption{Residual spectra of the light curve prewhitened for the solutions of Eqs.~\ref{eq:four} ($p$=185),  \ref{eq:summod} ($p$=185), and \ref{eq:benko} ($p$=55) are  shown. Note that the y-scales are the same in the three panels. The vertical grid indicates the positions of the pulsation frequency and its harmonics.  }
\label{sp}
\end{centering}
\end{figure}

Table \ref{rms} compares the rms of the residuals of several light-curve solutions using Eqs.~\ref{eq:four}, \ref{eq:summod} and \ref{eq:benko} with different parameter combinations.
In Case A and B, the solutions that have the same number of parameters have  similar accuracies; the rms of the Fourier solution (Eq.~\ref{eq:four}) is only marginally, $3-5$ per cent  smaller  than the rms of the amplitude and phase modulation formalism according to Eq.~\ref{eq:summod}. This is not at all the case, however, when using Eq.~\ref{eq:benko}. Even if similar number of parameters are fitted (57/55, 39/43), the residuals of Eq.~\ref{eq:benko} are $60-90$ per cent larger than the residuals of the other solutions. This situation cannot be improved by increasing the number of parameters when fitting Eq.~\ref{eq:benko}, as the rms of the fit decreases only marginally (by 0.5 per cent) when the orders of
the pulsation and the modulation are increased from (16, 2) to (18, 4). If even higher orders (20, 5) of the pulsation and the modulation are fitted the decrement of the residual is only 0.3 per cent. We note here, however, that fitting fifth-order amplitude and phase modulations is quite unrealistic, as even in the case of the very complex modulation of V445 Lyr, only a third-order FM (frequency modulation) and a first order AM (amplitude modulation) were assumed \citep{gug}. Also as neither pulsation nor modulation components have been detected in the Fourier spectrum at harmonic orders higher than 18, the  $n$=20 solution has no relevance compared to the $n$=18 one. As no secondary modulation of CM UMa has been detected, therefore the parameter number when applying the formalism of Eq. 7 cannot be increased this way either. Therefore, the $p$=55 ($n$=18, $N$=4) solution of the
BSP11 formalism is accepted and used for the comparison.

The similar accuracy of the solutions of Eqs.~\ref{eq:four} and \ref{eq:summod}  confirms that these forms are, in fact, equivalent in practice. 
Of course, minor difference between these solutions may
arise from the different role and significance of the parameters in the two formulae.  Therefore, the solutions do not necessarily have to have exactly the same accuracies
when the same number of parameters ($n,N,l$) are fitted.

In the top panels of Fig.~\ref{fit}, the light curve in the lowest and highest amplitude phases of the modulation are shown and their fits using Eq.~\ref{eq:four} ($p$=185), Eq.~\ref{eq:summod} ($p$=185) and Eq.~\ref{eq:benko} ($p$=55) are overplotted. (The $p$=63 solution of Eq.~\ref{eq:benko} does not give any noticeable difference if compared to the results of the  $p$=55 one.)
While the first two fits are indistinguishable from each other and they follow even the smallest features of the observed light variations accurately, this is not true for the fit obtained using 
Eq.~\ref{eq:benko}. Large, $0.02-0.05$~mag systematic differences, especially around light minima and maxima, appear between this fit and the observations. The bottom panels show the differences (fit -- observation) for the three light-curve solutions.  While the first two show nothing but noise around zero, in the third case, the fit does not follow the light variation correctly in any phase of the pulsation.

The spectra of the three residuals shown in Fig.~\ref{sp} also document that the  differences between the results are substantial. While the mean levels of the residual spectra are around 0.0003 mag in the first two cases, it is as high as 0.0009 mag in the third one. Moreover,
when using Eq.~\ref{eq:benko}, $0.006-0.009$~mag amplitude signals at around the positions of the pulsation frequency and its harmonics remain in the residual.

\subsection{Comparison of the amplitude and phase relations of the fits}
\label{a-p}

\begin{figure}
\includegraphics[width=9 cm]{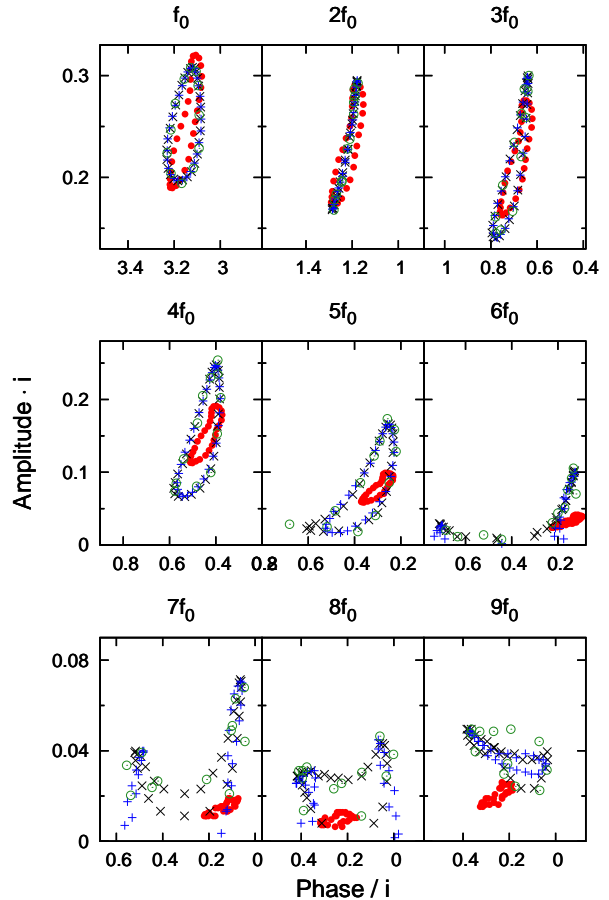}
\begin{centering}
\caption{The observed and fitted amplitudes vs phases of the different harmonic orders of the pulsation light curve during the Blazhko cycle are shown. The phase values are divided and the amplitudes are multiplied by the number of the harmonic order ($i$) in each plot for a better visualisation and comparison.
Note that the same, 0.7-rad phase ranges are shown in each panel, while the amplitude ranges are different in the different rows of panels. The symbols denoting the observations and the three different synthetic fits are the same as in Fig.~\ref{fit}. }
\label{egg}
\end{centering}
\end{figure}

The maximum phase\,--\,maximum brightness diagrams of Blazhko RR Lyrae stars show a large variety. The direction of going around these loops and their shape are defined by the relative amplitudes and  phases of the maximum-brightness and maximum-phase variations during the Blazhko cycle. The same is true for the shapes of the maximum phase\,--\,maximum brightness loops in the different harmonic orders of the pulsation. These curves can be determined from the amplitude and phase variations of the different harmonics of the pulsation light curve during the Blazhko cycle. 

The amplitude and phase variations have been shown to have different shapes in the different harmonic orders even in a very regular Blazhko star, MW Lyr \citep[see figure 12 in][]{mw}.
Striking differences between the variations of the amplitudes of $f_0$ and $2f_0$ ($A_1$ and $R_{21}$) of RR Lyr have been also shown \citep[see figure 12 in][]{rrl}.
 %In strongly irregular cases, the differences between e.g., the phase variations of the harmonic order components can differ from one Blazhko cycle to the other \citep[see figure 5 in][]{gu11}.

In this section, the amplitude and phase variations in the different harmonic orders of the light curve of CM~UMa are shown, and compared to the same plots derived from synthetic data according to the three light-curve solutions introduced in the previous section.

Using the parameters of the $p$=185, $p$=185 and $p$=55 solutions of the three different formalisms, synthetic light curves of CM~UMa have been generated. Dividing the synthetic data into small, homogeneous Blazhko phase bins, the variations of the pulsation light curves during the Blazhko cycle predicted by the different fits can be followed and compared to the observations. Both the observed and the fitted pulsation light curves in the different Blazhko phases are then fitted with a 15th-order Fourier sum of the pulsation frequency. The phases and amplitudes of the $f_0, 2f_0,..if_0$ components characterize the modulations of the pulsation components in the different orders; they define the maximum phase--maximum brightness loops. 

In Fig.~\ref{egg}, the amplitude vs phase variations of the first nine harmonics of the pulsation during the Blazhko cycle are plotted. The plotted data are derived from 12 and 30 Blazhko-phase bins of the observations and the fits, respectively. The phase variations are normalized by the division by the harmonic order $i$ (cf. Eq.~\ref{eq:benko}) and the amplitude variations are multiplied by the harmonic order for better visualisation.
The synthetic data using Eqs.~\ref{eq:four} and ~\ref{eq:summod} fit the observations in the first six orders absolute correctly. In the 7-9th orders, minor differences both between these fits and the fits and the observed values appear, because the observed values are somewhat uncertain in these harmonic orders. 

The observed and fitted (via Eqs.~\ref{eq:four} and ~\ref{eq:summod}) loops 
are significantly different in the different orders; the most striking differences are in the phase ranges of the loops. The phase ranges, even if normalized by the division of the order number vary between 0.1 and 0.6 rad. Large diversity in the morphology of the loops are also evident, e.g. it is `egg'-shaped in the first order while in the 2nd and 6th orders the loops are degenerated, they are reduced to one-dimensional curves in these harmonic orders as the amplitude and phase variations (see Fig.~\ref{egg}) are quite symmetrical in these orders, and their phase differences are $180\degr$.

The synthetic data corresponding to Eq.~\ref{eq:benko} do not follow the real variations of the phases and amplitudes of the maxima with the harmonic orders, due to the strong restrictions of the variations in the amplitude and phase modulations of the formalism of BSP11. The morphology of these loops are identical in each order, as differences only in the relative scaling of the amplitudes are allowed.

\begin{figure}
\includegraphics[width=8.3 cm]{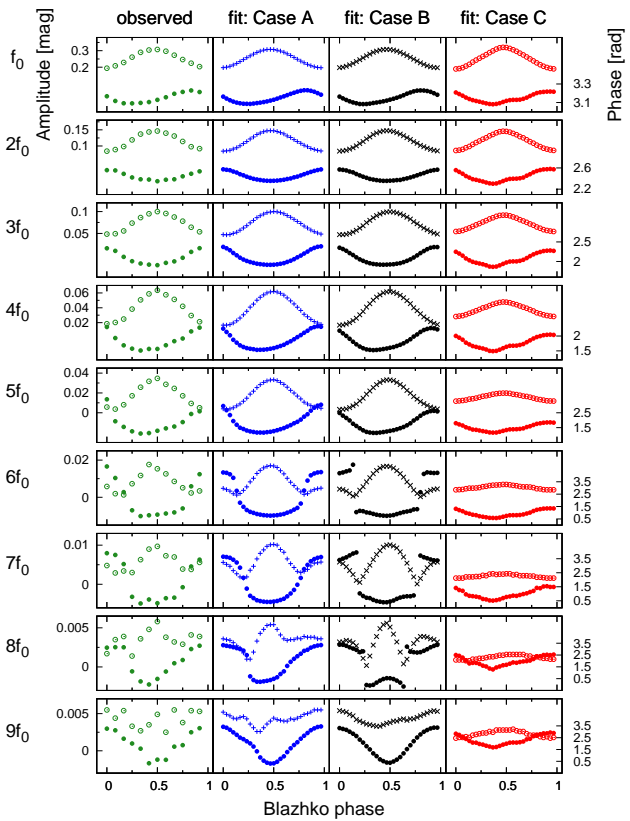}
\begin{centering}
\caption{Variations of the Fourier amplitudes and phases of the harmonic orders of the pulsation light curves during the Blazhko cycle. From the left to the right, the panels show the results obtained directly from the observations and from the three different fits. Filled symbols denote the phase variations.}
\label{four}
\end{centering}
\end{figure}

Fig.~\ref{four} documents the phase and amplitude variations of the pulsation light curve versus Blazhko phase in the different harmonic orders. The observations and the results of the three fits are plotted in separate columns here. 
Different amplitude and phase ranges are shown for the different harmonic orders but the same ranges are used for the observed and fitted values in each order (in each row in Fig.~\ref{four}).  Again, we can see that the first two fits give back the observed variations similarly well, with high precision. In Case C, however, neither the mean values nor the amplitudes and shapes of the predicted variations match the observations correctly. The  BSP11 formalism is not capable of fitting the observed differences of the phase relations between the amplitude and phase variations in the different orders.

\section{Conclusions }

The Fourier sequence describing the light curve of a Blazhko star (Eq.~\ref{eq:four}) has been transformed into the form of an amplitude and angle modulated signal (Eq.~\ref{eq:summod}). The two descriptions are fully equivalent to each other. The correctly deduced amplitude and angle modulation functions, $f_{\mathrm{A}i}(t)$ and $f_{\mathrm{F}i}(t)$  depend upon the harmonics of the unmodulated light curve ($i$). Since these are periodic functions, they can be expressed as Fourier sums (Eqs.~\ref{amod} and \ref{pmod}). A priori, there is no knowledge about the parameters of these equations, they can be determined only through observations. The possible correlations among them can be revealed by observations, as well.

In their study, BSP11 employed an amplitude and angle modulation pattern used in electronic signal transmission. (In this technique, the coding of the modulation of a signal is known in advance and can be controlled.) They assume that the amplitude modulation function does not depend upon the harmonic order of the unmodulated signal (light curve) and the difference between the angle modulation functions in the different orders is simply the multiplication by the harmonic-order number. This oversimplified procedure leads to the drastically reduced number of parameters but, on the cost of an unacceptably poor fit.

The analysis of the light curve of the Blazhko RR Lyrae star, CM~UMa (Sect.~\ref{cmu}), clearly shows that the BSP11 formalism gives a poor fit, especially around the minimum and maximum light and the ascending branch of the light curve. Even if the number of the parameters are increased, the fit does not improve. It should be emphasized that these parts of the light curve reflect the changing strength and occurrence of the shock-waves during the Blazhko cycle \citep{psp,cg97}. 
We thus conclude that the BSP11 approach does not take the nonlinear interactions, which determine the final form of the light variation, fully into account because of the strong restrictions of the formalism. Its capability to describe the changes of the pulsation light-curve's shape is seriously limited, since it only shifts the pulsation light curve in phase and scales it in amplitude periodically during the modulation.

Plots that show the amplitude and phase variations of the different harmonic orders during the Blazhko cycle (Figs.~\ref{egg} and \ref{four})  reveal it convincingly that the amplitude and angle modulation functions given in BSP11 are unsuited to describe the Blazhko modulation correctly.

Nevertheless, the Blazhko modulation can be correctly interpreted as amplitude- and angle-modulated signal (Eqs.~\ref{eq:summod}--\ref{pmod}). This fact hints at the possibility that during the Blazhko cycle the radial pulsation of an RR Lyrae star is modulated in phase/frequency and amplitude, and the fundamental period is subject to real oscillation. Up to now, the only model that is in conformation with this scenario has been suggested by \cite{st1}, although the inside physics of this model has been  strongly criticized by \cite{sm} and \cite{mol} recently.

\section*{Acknowledgments}

The financial support of OTKA grant K-81373 is acknowledged.

\appendix

\section{Trigonometric identities}

From the well-known identities of trigonometry, we obtain:

\begin{equation}
\label{eq:sinid}
A_1 \sin \left( \omega + \chi_1 \right) + A_2 \sin \left( \omega + \chi_2 \right) = A \sin \left( \omega + \chi \right)
\end{equation}

where $$A=\sqrt{A_1^2 +A_2^2 +2A_1A_2 \cos \left( \chi_2 - \chi_1 \right)} $$ and $$\tan \chi = \frac{A_1 \sin\chi_1 + A_2\sin\chi_2}{A_1\cos\chi_1+A_2\cos\chi_2}$$

By reiterating the formula \ref{eq:sinid}, we come to the relation

\begin{equation}
\label{eq:soksin}
\sum_{i=1}^n A_i\sin\left( \omega + \chi_i\right) = A \sin\left( \omega + \chi \right)
\end{equation}

where both $A$ and $\chi$ are bounded functions of $(A_1, A_2, ..., A_n, \chi_1, \chi_2, ..., \chi_n)$. This relation holds even if $A_i$ and $\chi_i$ are time dependent.

\section{Frequency and phase modulations}

Two types of angle modulation exist, frequency (FM) and phase (PM) modulations.

In the case of FM, the frequency of the unmodulated wave (carrier signal), $f_0$, is modulated by the modulating signal,
$\Theta_\mathrm{F}(t)$. The instantaneous frequency is

\begin{equation}
\label{eq:b_instfreq}
 F(t) = f_0 + k_\mathrm{F} \Theta_\mathrm{F}(t),
\end{equation}

\noindent where $k_\mathrm{F}$ constant depends upon the modulating
system. The instantaneous phase of the modulated wave is

\begin{equation}
\label{eq:b_instphase}
 \Psi(t) = 2\pi f_0 t + 2\pi k_\mathrm{F}
\int_0^t\Theta_\mathrm{F}(\tau)\,\mathrm{d}\tau.
\end{equation}

\noindent (For simplicity, zero phase is assumed at $t=0$.)

In the case of PM, the phase of the modulated wave is

\begin{equation}
\label{eq:b_instphase2}
 \Psi(t) = 2\pi f_0 t + k_\mathrm{P} \Theta_\mathrm{P}(t),
\end{equation}

\noindent where $\Theta_\mathrm{P}(t)$ and $k_\mathrm{P}$ are the
modulating signal and a constant, respectively. The instantaneous
frequency of a PM wave is

\begin{equation}
\label{eq:b_instfreq2}
 F(t) = f_0 + k_\mathrm{P} \frac{\mathrm{d} \Theta_\mathrm{P}(t)}{\mathrm{d} t}.
\end{equation}

In both cases, the wave's frequency and phase vary from moment to
moment. From Eqs.~\ref{eq:b_instfreq} and \ref{eq:b_instfreq2} follows
that the two modulations describe the same modulated signal, if the
following relation holds:

\begin{equation}
 \label{b_modsignals}
 \Theta_\mathrm{F}(t) = \frac{k_\mathrm{P}}{k_\mathrm{F}}
\frac{\mathrm{d} \Theta_\mathrm{P}(t)}{\mathrm{d} t}.
\end{equation}

\noindent The only difference between the two descriptions is that in the case of FM, the frequency modulation of the carrier is given by the time derivative of the PM modulated signal. 
%Therefore, the PM is often considered as a special case of FM. 
Unless some information is available about the modulation in advance, (as it is the case in electronic signal transmission), it may not be obvious, which one of the two types (FM or PM) is realized. If no information is available regarding the angle modulation (as in the case of Blazhko modulation), it is impossible to identify it as an FM or PM signal. This explains why the expression `angle modulation' is used thorough this paper.

\label{lastpage}

\begin{thebibliography}{}
\bibitem[Alcock et al.(2003)]{alcock} Alcock C., et al.,  2003, ApJ, 43,217  
\bibitem[Benk\H o, Szab\'o \& Papar\'o(2011)]{b11}  Benk\H o J., Szab\'o R.,  Papar\'o M., 2011, MNRAS, 417, 974 (BSP11)
\bibitem[Blazhko(1907)]{bl} Blazhko S., 1907, AN, 175, 325
\bibitem[Breger \& Kolenberg(2006)]{bk} Breger M., Kolenberg K., 2006, A\&A 460, 167
\bibitem[Chadid(2009)]{ch09} Chadid M., 2009, in Guzik J. A., Bradley P. eds, AIP Conf. Proc. 1170, Stellar Pulsation: Challenges for Theory and Observation, p. 235
\bibitem[Chadid \& Gillet(1997)]{cg97} Chadid M., Gillet D., 1997, A\&A, 319, 154
\bibitem[Guggenberger et al. (2012)]{gug} Guggenberger E., et al., 2012, MNRAS accepted, 2012arXiv1205.1344G
\bibitem[Jurcsik et al.(2008a)]{mw}  Jurcsik J. et al., 2008a, MNRAS, 391, 164
\bibitem[Jurcsik et al.(2008b)]{mw2}  Jurcsik J. et al., 2008b, MNRAS, 393, 1553
\bibitem[Jurcsik et al.(2009)]{kbs} Jurcsik J., et al., 2009, MNRAS, 400, 1006	%blstat.
\bibitem[Kolenberg et al.(2010)]{k10} Kolenberg K., et al., 2010, ApJL, 713, 198
\bibitem[Kolenberg et al.(2011)]{rrl} Kolenberg K. et al., 2011, MNRAS, 411, 878 %rr lyr
\bibitem[Kolenberg(2011)]{k11} Kolenberg K., 2011, in: RR Lyrae Stars, Metal-Poor Stars, and the Galaxy, ed, A.M.McWilliam, Carnegie Observatories Astrophysics Series, Vol. 5, 100
\bibitem[Kov\'acs(2009)]{ko09} Kov\'acs G., 2009, in Guzik J. A., Bradley P. eds, AIP Conf. Proc. 1170, Stellar Pulsation: Challenges for Theory and Observation, p. 261
\bibitem[Moln\'ar, Koll\'ath \& Szab\'o(2012)]{mol} Moln\'ar L.,  Koll\'ath Z. \& Szab\'o R, 2012 MNRAS in press, 2012arXiv1203.2911M
\bibitem[Preston, Smak \& Paczy\`nski(1965)]{psp} Preston G. W., Smak J., \& Paczy\`nski B., 1965, ApJS, 12, \bibitem[\protect\citeauthoryear{S\'odor}{2012}]{kbs2}  S\'odor \'A., et al. 2012, in: proceedings of the 61st Fujihara Seminar, ed.: Hiromoto
Shibahashi, ASPC, 2012
 \bibitem[Shapley(1916)]{sh} Shapley H., 1916, ApJ, 43, 217
\bibitem[\protect\citeauthoryear{Smolec et al.}{2011}]{sm} Smolec R., Moskalik P., Kolenberg K., Bryson S., Cote M. T., Morris R. L., 2011, MNRAS, 414, 2950
\bibitem[Soszy\`nski et al.(2003)]{scz03} Soszy\`nski I., et al., 2003 Acta Astr. 53, 93
\bibitem[Soszy\`nski et al.(2011)]{scz11} Soszy\`nski I., et al., 2011 Acta Astr. 61, 1
\bibitem[Stothers(2006)]{st1} Stothers R., 2006, ApJ, 652, 643
\bibitem[Szeidl \& Jurcsik(2009)]{szj} Szeidl B., Jurcsik J., 2009 CoAst 160, 17
\end{thebibliography}
\end{document}